\documentclass[a4paper,12pt]{article}

\usepackage[bf]{titlesec}
\usepackage[textwidth=16.5cm,textheight=25cm]{geometry}
\usepackage[margin=27pt,font=small,labelfont=bf]{caption}
\usepackage{fancyhdr}
\usepackage{amsmath}
\usepackage{amssymb}
\usepackage{amsfonts}
\usepackage{color}
\usepackage{graphicx}
\usepackage{array}
\usepackage[nofiglist,notablist,nomarkers]{endfloat}

\usepackage{tikz}

\usepackage{subcaption}
\newlength\fwidth
\newlength\fheight

\newcommand{\mytitle}{\large Spermatozoa scattering by a microchannel feature: an elastohydrodynamic model}
\newcommand{\myauthors}{\normalsize T. D. Montenegro-Johnson\textsuperscript{1,2,3}\footnote{Author for correspondence, email: \texttt{tdj23@cam.ac.uk}}, H. Gad\^{e}lha\textsuperscript{4,3} and D. J. Smith\textsuperscript{2,3,5}.}

\titleformat*{\section}{\large\bfseries}
\titleformat*{\subsection}{\slshape}

\begin{document}

\title{\textbf{\mytitle}}
\author{\myauthors}
\date{}

\maketitle

\begin{center}
\begin{tabular}{p{0.01\textwidth}p{0.84\textwidth}}
~~$^1$ &Department of Applied Mathematics and Theoretical Physics, Centre for Mathematical Sciences, University of Cambridge, Wilberforce Road, Cambridge, CB3 0WA, U.K. \\[1mm]
~~$^2$ &School of Mathematics, University of Birmingham, Edgbaston, Birmingham, B15 2TT, U.K.\\[1mm]
~~$^3$ &Centre for Human Reproductive Science, Birmingham Women's NHS Foundation Trust, Mindelsohn Way, Edgbaston, Birmingham,  B15 2TG, U.K. \\[1mm]
~~$^4$ &Wolfson Centre for Mathematical Biology, University of Oxford, Mathematical Institute, Woodstock Road, OX2 6GG, U.K. \\[1mm]
~~$^5$ &School of Engineering and Centre for Scientific Computing, University of Warwick, Coventry, CV4 7AL, U.K.
\end{tabular}
\end{center}

\vspace*{0.5in}

\begin{abstract}
Sperm traverse their microenvironment through viscous fluid by propagating
flagellar waves; the waveform emerges as a consequence of elastic structure,
internal active moments, and low Reynolds number fluid dynamics. Engineered
microchannels have recently been proposed as a method of sorting and
manipulating motile cells; the interaction of cells with these artificial
environments therefore warrants investigation. A numerical method is presented
for large-amplitude elastohydrodynamic interaction of active
swimmers with domain features. This method is employed to examine hydrodynamic
scattering by a model microchannel backstep feature. Scattering is shown to
depend on backstep height and the relative strength of viscous and elastic
forces in the flagellum. In a `high viscosity' parameter regime corresponding to
human sperm in cervical mucus analogue, this hydrodynamic contribution to
scattering is comparable in magnitude to recent data on contact effects, being
of the order of $5$--$10^\circ$. Scattering can be positive or negative
depending on the relative strength of viscous and elastic effects, emphasising
the importance of viscosity on the interaction of sperm with their
microenvironment. The modulation of scattering angle by viscosity is associated
with variations in flagellar asymmetry induced by the elastohydrodynamic
interaction with the boundary feature.
\end{abstract}

\begin{center}
Key index words: \emph{Stokesian swimming, fluid-structure interaction, human sperm}
\end{center}

\section{Introduction}

Human sperm propel themselves by propagating a travelling wave along a single,
active flagellum; this motility is essential for migration through the female
reproductive tract and natural fertilisation. Recent work with microfluidic
devices \cite{denissenko2012human,kantsler2013ciliary} has suggested the ability
to direct and sort cells through their own motility, a potentially valuable
advance in assisted reproduction therapy and in the livestock industry. Cell
scattering at simple geometric features, such as the outside of a corner, appear
to be dependent on viscosity and temperature; developing mechanical models to
understand, interpret and optimise these effects for their exploitation is
therefore of considerable interest. We will develop a mathematical model of a
cell interacting with its environment, and its computational implementation, and
study the dynamics of a realistic model sperm swimming over a backstep feature
to study the effect of elastic, viscous and geometric parameters. The model will
combine geometric nonlinearity of the elastic flagellum with nonlocal
hydrodynamic interactions, and will be solved numerically via an implicit finite
difference method for the elastohydrodynamic equations, combined with a
hybrid slender body theory/boundary integral method for
the low Reynolds number fluid dynamics.

The motor apparatus driving the flagellar waveform is a remarkably
phylogenetically conserved structure known as the axoneme. The axoneme in human
sperm comprises $9$ doublet microtubules, linked to each other and a central
pair by passive elastic structures, with additional stiffening from outer dense
fibres and the fibrous sheath (for recent review focused on
mechanically-relevant features, see ref.~\cite{gaffney2011}). Motor proteins
bound to microtubules exert forces on adjacent doublets in a coordinated manner
to induce bending moments along the length of the flagellum, causing bending,
which is in turn resisted by the surrounding fluid. The fluid mediates
interactions with surrounding surfaces and other cells; the flagellar waveform
emerges from this nonlinear coupling.

Machin~\cite{machin1958wave} showed that in order to generate experimentally
observed waveforms, the flagellum must actively bend along its length,
developing a linearised theory that has formed the basis of many subsequent
studies. The theory that bending is produced by relative sliding of internal
microtubules was subsequently proposed by Satir~\cite{satir1965studies}, and the
sliding mechanism was modelled in early studies by
Brokaw~\cite{brokaw1971bend,brokaw1972computer}, using the formalism of an
active internal moment per unit length in an elastic filament. The regulation of
the active motor proteins that cause this sliding, and their oscillatory
behaviour, is however a subject of continuing
enquiry~\cite{brokaw2009,lindemann2010,woolley2010}, with modelling playing an
important role in comparing regulatory theories \cite{riedel2007molecular}. A
number of studies since the 1970s have provided significant insights into how
potential mechanisms of dynein regulation can produce the types of bending waves
observed in nature (see for
example~\cite{lindemann1994geometric,hines1978bend,camalet2000generic,brokaw2002,brokaw2009}).

The importance of large-amplitude elastohydrodynamic flagellar
modelling  was established by Gad\^{e}lha et al.~\cite{gadelha2010nonlinear},
who delineated the range of validity of small-amplitude
elastic theory and showed that for sufficiently high viscosity relative to
flagellar stiffness, a buckling instability can give rise to waveform asymmetry
without domain boundaries or asymmetric internal actuation. The numerical
implementation of Gad\^{e}lha et al.'s study built on a
model of passive flexible fibres in shear flow~\cite{tornberg2004simulating},
although replacing the nonlocal hydrodynamics of the latter with a local
drag-velocity law. The combination of three-dimensional, time-dependent flow
with the hydrodynamic interactions arising from fixed and moving boundaries,
with active filament mechanics is computationally
demanding; the majority of sperm models until the last decade made similar
approximations for the fluid dynamics, or small-amplitude linearisation of the
flagellar wave.

Liron, Gueron and colleagues (see for example~\cite{gueron1992,liron2001lgl})
modelled cilia arrays, taking both nonlocal fluid dynamics and geometric
nonlinearity into account, building on earlier work by for example
Lighthill~\cite{lighthill1976flagellar} and Hines \& Blum~\cite{hines1978bend}.
However this formalism, expressed in terms of bending angles rather than
flagellar position, does not appear to have been generalised to a free-swimming
cell with the associated boundary condition resulting from the presence of a
head. More recent work using the finite element and finite volume methods and
cluster computing has also been focused on cilia~\cite{mitran2007}; another
successful recent approach is the regularised stokeslet method combined with a
generalised immersed boundary method~\cite{olson2013modeling}.

While the fluid dynamic interaction of sperm with plane boundaries has received
significant attention since the work of Rothschild over 50 years
ago~\cite{rothschild1963non}, motivating a number of experimental and
theoretical studies \cite{winet1984,fauci1995,smith2009human,elgeti2010}, the
interaction of sperm with `non-trivial' geometric obstacles involving angles and
curves or complex interfaces is a subject of growing recent interest
\cite{crowdy2011,davis2012,crowdy2013,lopez2014}.

Denissenko et al.~\cite{denissenko2012human} showed how sperm scatter at a range
of angles when encountering the outside of a corner in an artificial
microchannel maze, and that the scattering angle is modulated by viscosity;
Kantsler et al.~\cite{kantsler2013ciliary} studied the effect of very close
interactions of sperm and the biflagellate algae \emph{Chlamydomonas} with these
features.  The geometric nature of the female reproductive tract is also highly
convoluted, further motivating the need for models which can accommodated
complex wall shapes into account. These studies suggest tantalising
opportunities to direct and sort motile sperm on passive microdevices, however a
better understanding of the subtle nonlinear physics of how flagellated swimmers
interact with geometric features must be developed; to aid with this
understanding we will develop a mathematical and computational approach which
accounts for elasticity, viscosity and their interaction, without the need for
large scale computational resources. To this end we will bring together the
active elastic formulation of Gad\^{e}lha \textit{et al.}
\cite{gadelha2010nonlinear} with the Lighthill-Gueron-Liron theorem
\cite{gueron1992} for nonlocal slender body theory and the boundary element
\cite{pozrikidis2010} and regularised stokeslet methods
\cite{cortez2005method,smith2009boundary} to capture the influence of a
non-trivial nearby surface. We will use this approach to explore how sperm
scatter near geometric features due to elastohydrodynamic interaction over
hundreds of flagellar beats with a single computer core, and quantify how the
balance of viscosity and elasticity modulates this effect via changes to the
flagellar waveform.

\section{Mathematical model}

The mathematical model of a sperm interacting with a geometric feature will be
derived from, (i) the Stokes flow equations, with a nonlocal hydrodynamic model,
and (ii) geometrically nonlinear elasticity for an internally actuated
flagellum. We will first derive the equations for the two parts of problem,
before describing (iii) the numerical implementation.

\subsection{Hydrodynamics}

At microscopic scales, fluid dynamics can be modelled by the incompressible Stokes flow equations,
\begin{equation}
0=-\boldsymbol{\nabla}p+\mu\nabla^2\mathbf{u} \mbox{,} \quad
\boldsymbol{\nabla}\cdot\mathbf{u} = 0,
\end{equation}
where $\mathbf{u}$ is velocity, $p$ is pressure and $\mu$ is dynamic viscosity.
For our problem, these equations will be augmented with the no-slip,
no-penetration condition $\mathbf{u}(\mathbf{X})=\mathbf{X}_t$ for points
$\mathbf{X}$ on the solid boundary, where subscript $t$ denotes time derivative.

The linearity of the Stokes flow equations enables the construction of solutions
to satisfy boundary conditions via discrete and/or continuous sums of
suitably-weighted fundamental solutions. These techniques replace solid
surfaces, such as the sperm flagellum, head, and its surrounding
microenvironment, by line or surface distributions of immersed forces. A
concentrated point force located at $\mathbf{y}$ with strength $\mathbf{F}$,
produces a velocity field (the `stokeslet'),
\begin{equation}
u_j(\mathbf{x})=S_{jk}(\mathbf{x},\mathbf{y})F_k \mbox{,} \quad \mbox{where} \quad S_{jk}(\mathbf{x},\mathbf{y})=\frac{1}{8\pi\mu}\left(\frac{\delta_{jk}}{|\mathbf{x}-\mathbf{y}|}+\frac{(x_j-y_j)(x_k-y_k)}{|\mathbf{x}-\mathbf{y}|^3}\right)\mbox{,}\label{eq:stokeslet}
\end{equation}
the symbol $\delta_{jk}$ being the Kronecker delta tensor and the summation
convention being used. The symbol
$\boldsymbol{\mathsf{S}}(\mathbf{x},\mathbf{y})$ will be used to denote the 2nd
rank tensor in equation~\eqref{eq:stokeslet}. It will
also be convenient to make use of the regularised stokeslet
$\boldsymbol{\mathsf{S}}^\epsilon$ of Cortez~\cite{cortez2001method}, which
corresponds to a spatially smoothed force; a frequently-used implementation in
three dimensions \cite{cortez2005method} takes the form,
\begin{equation}
S_{jk}^\epsilon(\mathbf{x},\mathbf{y})=\frac{1}{8\pi\mu}\frac{\delta_{jk}(|\mathbf{x}-\mathbf{y}|^2+2\epsilon^2)+(x_j-y_j)(x_k-y_k)}{(|\mathbf{x}-\mathbf{y}|^2+\epsilon^2)^{3/2}}.
\end{equation}
The parameter $\epsilon>0$ defines the length scale over which the
point force is smoothed; this smoothness property is particularly convenient for the formulation of boundary integral methods.

The LGL theorem \cite{liron2001lgl}, an extension of the work of
Lighthill~\cite{lighthill1976flagellar}, derives from a line distribution of
singular stokeslets and source dipoles: an approximate expression for the flow
field at the surface of a moving slender body, accurate to $O(\sqrt{b/L})$,
where $b$ is the radius and $L$ is the flagellar length. Ignoring image systems,
which are not required in our formulation, and using the properties of the
stokeslet to reorder the source and field points, we have the expression for the
approximate velocity field produced by the slender body $\mathbf{v}$,
\begin{align}
\mathbf{v}(\mathbf{X}(s_0,t)) &= -\frac{1}{\xi_{\parallel}}(\mathbf{f}_{\mathrm{vis}}\cdot \hat{\mathbf{s}})\hat{\mathbf{s}}-\frac{1}{\xi_{\perp}}(\mathbf{f}_{\mathrm{vis}}\cdot \hat{\mathbf{n}})\hat{\mathbf{n}}-\frac{1}{\xi_{\perp}}(\mathbf{f}_{\mathrm{vis}}\cdot \hat{\mathbf{b}})\hat{\mathbf{b}}\nonumber \\
& -\int_{|s-s_0|>q} \boldsymbol{\mathsf{S}}(\mathbf{X}(s_0,t),\mathbf{X}(s,t))\cdot \mathbf{f}_{\mathrm{vis}}(s)\,ds.\label{eq:lgl}
\end{align}
Here and in what follows, $0\leqslant s \leqslant L$ is an arclength
parameterisation for the flagellum, and $\mathbf{f}_{\mathrm{vis}}$ is the
viscous force per unit length exerted by the fluid on the flagellum. The
coefficients $\xi_{\parallel}$ and $\xi_{\perp}$ are parallel and perpendicular
resistance coefficients similar to those of Gray \&
Hancock~\cite{gray1955propulsion} and take the form,
\begin{equation}
\xi_{\perp} = \frac{8\pi\mu}{1 + 2 \ln (2q/b)}, \quad \xi_{\parallel} =
\frac{8\pi\mu}{-2 + 4\ln (2q/b)}, \quad \gamma = \frac{\xi_{\perp}}{\xi_{\parallel}}\mbox{,}
\end{equation}
the parameter $q$ being a length scale chosen
intermediate in magnitude between $b$ and $L$. The symbols $\hat{\mathbf{s}}$,
$\hat{\mathbf{n}}$ and $\hat{\mathbf{b}}$ are unit tangent, normal and binormal.
Whereas Gueron and Liron~\cite{gueron1992,liron2001lgl} considered the dynamics
of a cilium projecting from a plane boundary, and hence the associated image
systems, in this study we will not require these terms because surfaces will be
represented via boundary integrals.

Equation~\eqref{eq:lgl} can be considered a nonlocal extension of resistive
force theories, which retain only the first three terms. To couple LGL to the
elastohydrodynamic model of Gad\^{e}lha et al.~\cite{gadelha2010nonlinear} we
will rewrite these terms in another commonly-used form,
$-(1/\xi_{\perp})(\boldsymbol{\mathbf{I}}+(\gamma-1)\hat{\mathbf{s}}\hat{\mathbf{s}})\cdot
\mathbf{f}_{\mathrm{vis}}$, with $\gamma=\xi_{\perp}/\xi_{\parallel}$ playing a
similar role to the drag anisotropy ratio of resistive force theory, but depending
on the choice of $q$. The precise value of $q$ is not critical provided that $b\ll q \ll L$ because changes to the resistance coefficients are accompanied by changes to the integrals; for our study with $b=0.01L$, we choose $q=0.1L$, leading to $\gamma\approx 1.4$.

To model a sperm, we will consider a cell with a rigid head as well as a
flagellum, swimming near a rigid step-like surface. The linearity of Stokes flow
equations means that a solution satisfying the additional no-slip boundary
conditions associated with the head and the wall may be constructed by linear
superposition. Moreover, the Lorentz reciprocal relation, and its regularised
analogue \cite{cortez2005method} enable the representation of these surfaces by
boundary integrals; rigidity of the surfaces enables the use of single layer boundary integral
representations \cite[p. 32]{pozrikidis1992}. In the present study we will use a hybrid approach,
representing the head via a surface distribution of singular stokeslets with
stress $\boldsymbol{\phi}^{\mathrm{H}}$, discretised via BEMLIB
\cite{pozrikidis2010}, and the wall by regularised stokeslets and boundary
elements, with stress $\boldsymbol{\phi}^{\mathrm{W}}$
\cite{cortez2005method,smith2009boundary}. The full fluid dynamic model for the
velocity field on the surface of the flagellum is therefore,
\begin{align}
\mathbf{u}(\mathbf{X}(s_0,t))& = -\frac{1}{\xi_{\perp}}(\boldsymbol{\mathbf{I}}+(\gamma-1)\hat{\mathbf{s}}\hat{\mathbf{s}})\cdot \mathbf{f}_{\mathrm{vis}} -\int_{|s-s_0|>q} \boldsymbol{\mathsf{S}}(\mathbf{X}(s_0,t),\mathbf{X}(s,t))\cdot \mathbf{f}_{\mathrm{vis}}(s)\,ds \nonumber \\
&-\iint_{H(t)} \boldsymbol{\mathsf{S}}(\mathbf{X}(s_0,t),\mathbf{y})\cdot\boldsymbol{\phi}^{\mathrm{H}}(\mathbf{y})\,dS_{\mathbf{y}}-\iint_W \boldsymbol{\mathsf{S}}^\epsilon(\mathbf{y},\mathbf{X}(s_0,t))\cdot\boldsymbol{\phi}^{\mathrm{W}}(\mathbf{y})\,dS_{\mathbf{y}}.\label{eq:fluiddyn}
\end{align}
Similar equations, but with the first two terms replaced by a single slender
body integral
$-\int_0^L\boldsymbol{\mathsf{S}}\cdot\mathbf{f}_{\mathrm{vis}}\,ds$, hold on the
surface of the head and the wall. In the next section we will discuss the equations of an internally-driven
elastic flagellum, and their coupling to the fluid mechanics.

\subsection{Elastohydrodynamics}
The elastohydrodynamic formulation we will work with was derived by Tornberg \&
Shelley \cite{tornberg2004simulating}, and extended to an internally-driven
flagellum by Gad\^{e}lha et al.~\cite{gadelha2010nonlinear}; the central feature
of this approach is to formulate the problem in terms of the flagellar position
$\mathbf{X}(s,t)$ and line tension $T(s,t)$. Alternative approaches based on
bending angles and curvatures \cite{gueron2001,brokaw2001} have also been
pursued, as has complex curvature \cite{goldstein1998}. The internal elastic
contact force $\mathbf{F}_{\mathrm{int}}$ and moment $\mathbf{M}_{\mathrm{int}}$ exerted on the proximal flagellum $[0,s_0)$ by the
distal flagellum $(s_0,L)$, respectively are given by,
\begin{equation}
\mathbf{F}_{\mathrm{int}}=-E\mathbf{X}_{sss}+m\hat{\mathbf{n}}+T\mathbf{X}_s \mbox{,} \quad \mathbf{M}_{\mathrm{int}}\wedge \mathbf{X}_s = E\mathbf{X}_{ss} \mbox{,} \label{eq:elasticity}
\end{equation}
where $E$ is constant elastic modulus and $m(s,t)$ is a prescribed active moment
density representing the internal flagellar motors. Balancing elastic and viscous forces acting on a segment of flagellum $(s_0,s_0+\delta s)$
and taking the limit as $\delta s\rightarrow 0$ yields,
\begin{equation}
\mathbf{f}_{\mathrm{vis}} + \partial_s(-E\mathbf{X}_{sss}+m\hat{\mathbf{n}}+T\mathbf{X}_s) = 0.
\end{equation}
Noting that $\hat{\mathbf{s}}=\mathbf{X}_s$, the local term of
equation~\eqref{eq:fluiddyn} can then be written,
\begin{align}
-\frac{1}{\xi_{\perp}}(\boldsymbol{\mathbf{I}}+(\gamma-1)\hat{\mathbf{s}}\hat{\mathbf{s}})\cdot \mathbf{f}_{\mathrm{vis}}
&=
-E(\mathbf{X}_{ssss}+(\gamma-1)(\mathbf{X}_s\cdot\mathbf{X}_{ssss})\mathbf{X}_s)+T\mathbf{X}_{ss}+\gamma T_s\mathbf{X}_s\nonumber \\
& +m_s\hat{\mathbf{n}}+\gamma m \hat{\mathbf{n}}_s.
\end{align}
For brevity we will write the nonlocal (integral) velocities from
equation~\eqref{eq:fluiddyn} as $\mathbf{V}$ (written out explicitly in the appendix, equation~\eqref{eq:nonloc}). Nondimensionalising with scales
$L$ for position, $1/\omega$ for time, $\omega L$ for velocity and $E/L^2$ for
tension and moment density, yields the following dimensionless
elastohydrodynamic equation,

\begin{equation}
\mathrm{Sp}^4 (\mathbf{X}_t-\mathbf{V})  =
-\mathbf{X}_{ssss}-(\gamma-1)(\mathbf{X}_s\cdot \mathbf{X}_{ssss})\mathbf{X}_s +
T\mathbf{X}_{ss} + \gamma T_s \mathbf{X}_s  + m_s \hat{\mathbf{n}} + \gamma m \hat{\mathbf{n}}_s.\label{eq:elastohydro}
\end{equation}
The parameter $\mathrm{Sp}=L(\xi_{\perp}\omega/E)^{1/4}$ is the \emph{sperm
number}, which quantifies the relative importance of viscous and elastic
effects. This model can be seen as an extension of linear models
(such as Camalet et al.~\cite{camalet2000generic}) by the inclusion of the
nonlinear terms on the right hand side, and an extension of hydrodynamically
local models (such as Gad\^{e}lha et al.~\cite{gadelha2010nonlinear}) by the
inclusion of the $\mathbf{V}$ term on the left hand side.

Similarly to Gad\^{e}lha et al.~\cite{gadelha2010nonlinear}, the inextensibility
constraint $\partial_t(\mathbf{X}_s\cdot\mathbf{X}_s)=0$ can be used with the
elastohydrodynamic equation~\eqref{eq:elastohydro} to deduce an ordinary differential equation which
must be satisfied by the line tension $T$,
\begin{align}
-\mathrm{Sp}^4\mathbf{V}_s\cdot\mathbf{X}_s &=  \gamma T_{ss} - \mathbf{X}_{ss}\cdot\mathbf{X}_{ss}T + 3\gamma \mathbf{X}_{sss}\cdot\mathbf{X}_{sss}+(1+3\gamma)\mathbf{X}_{ss}\cdot\mathbf{X}_{ssss} \nonumber \\
& +(\gamma +1)m_s\hat{\mathbf{n}}_s\cdot\mathbf{X}_s + m\hat{\mathbf{n}}_{ss}\cdot\mathbf{X}_s.  \label{eq:aux}
\end{align}
The above equation is derived via the identity
$3\mathbf{X}_{ss}\cdot\mathbf{X}_{sss}+\mathbf{X}_s\cdot\mathbf{X}_{ssss}=0$ and
its derivative with respect to $s$. As previously
\cite{tornberg2004simulating,gadelha2010nonlinear} we introduce the term
$\lambda \mathrm{Sp}^4(1-\mathbf{X}_s\cdot\mathbf{X}_s)$ to the left hand side
of equation~\eqref{eq:aux} to dampen numerical errors in flagellar length. The
value used in the present study is $\lambda = 80$, though as found by Gad\^{e}lha
\textit{et al.} the solution is insensitive to the precise value of $\lambda$.

The final part of the mathematical model is the specification of the boundary
conditions for equations~\eqref{eq:elastohydro} and \eqref{eq:aux}. The
assumption of zero contact force and moment at the distal ($s=1$) tip of the
flagellum combined with the elasticity equations~\eqref{eq:elasticity} yield (in
dimensionless variables),
\begin{equation}
0=-\mathbf{X}_{sss}+m\hat{\mathbf{n}}+T\mathbf{X}_s \mbox{,}\quad 0=\mathbf{X}_{ss} \quad \mbox{at} \; s=1. \label{eq:distbc}
\end{equation}
Taking the dot product of the first equation with $\mathbf{X}_s$, using the
identity
$\mathbf{X}_s\cdot\mathbf{X}_{sss}=-\mathbf{X}_{ss}\cdot\mathbf{X}_{ss}$ and the
second equation yields the distal tension boundary condition, $T=0$.

At the proximal end of the flagellum, the boundary conditions are given by considering the force and moment exerted by the fluid on the head. We denote these quantities $\mathbf{F}^\mathrm{H}$ and $\mathbf{M}^\mathrm{H}$ and nondimensionalise them with the elastic scalings $E/L^2$ and $E/L$ respectively. In the inertialess Stokes flow regime, the total force and moment acting on the head are zero, so by Newton's third law, the force and moment on the flagellum at $s=0$ are also given by $\mathbf{F}^\mathrm{H}$ and $\mathbf{M}^\mathrm{H}$ respectively. With the appropriate scalings, the proximal boundary conditions are then,
\begin{equation}
\mathbf{F}^\mathrm{H}=\mathbf{X}_{sss}-m\hat{\mathbf{n}}-T\mathbf{X}_s \quad \mbox{and} \quad \mathbf{M}^\mathrm{H}\wedge\mathbf{X}_s=-\mathbf{X}_{ss}+M\hat{\mathbf{n}} \mbox{,} \quad \mbox{at} \; s=0 \mbox{,} \label{eq:proxbc}
\end{equation}
where $M=\int_0^1 m \,ds$. From these equations we also derive the tension
condition at the proximal end, $\mathbf{F}^\mathrm{H}\cdot \mathbf{X}_s =
-\mathbf{X}_{ss}\cdot \mathbf{X}_{ss} -T$. The calculation of the quantities $\mathbf{F}^\mathrm{H}$ and $\mathbf{M}^\mathbf{H}$ with nonlocal hydrodynamic interaction is described in more detail in the next section and the appendix. Finally we introduce the translational and angular velocity $\mathbf{U}^\mathrm{H}$ and $\boldsymbol{\Omega}^\mathrm{H}$ of
the head; while $\mathbf{U}^\mathrm{H}$ and two components of the angular
velocity are constrained by knowledge of the function $\mathbf{X}$, there is an
independent rotational component of the motion that defines the principal
bending plane of the flagellum. These quantities will be determined by kinematic
considerations and the implementation of the boundary conditions.

To complete the mathematical model it is necessary to specify the internal active moment $m(s,t)$. Gad\^{e}lha et al.~\cite{gadelha2010nonlinear} used travelling waves of internal moment, which calculations from experiment \cite{gaffney2011} confirm are a good model. We therefore specify in dimensionless units, $m(s,t)=m_0\cos(ks- t)$.

\subsection{Numerical implementation}
The elastohydrodynamic equation~\eqref{eq:elastohydro} is treated with a
Crank-Nicolson type finite difference discretisation, with the second order
central differences in the interior, and third order one-sided difference for
the boundary conditions, using coefficients taken from
Fornberg~\cite{fornberg1988generation}. The higher-order boundary stencil
produced comparable errors to the central stencil on polynomial test functions.
Both linear and nonlinear terms are treated implicitly; nonlinearity of these equations is dealt with by performing an iterative process on every timestep, with the operator on the left hand side at $t+dt$
being linearised as,
\begin{equation}
-\mathbf{X}_{ssss}-(\gamma-1)(\tilde{\mathbf{X}}_s\cdot\mathbf{X}_{ssss})\tilde{\mathbf{X}}_s+T\tilde{\mathbf{X}}_{ss}+\gamma
T_s\tilde{\mathbf{X}}_s+m_s\tilde{\mathbf{n}}+\gamma m\tilde{\mathbf{n}}_s \mbox{,}
\end{equation}
variables with tildes denoting that values from the previous iteration are
taken.

The nonlocal hydrodynamic term $\mathbf{V}$ in equation~\eqref{eq:elastohydro} is approximated by forming the slender
body/boundary integral problem of determining $\mathbf{f}_{\mathrm{vis}}$,
$\boldsymbol{\phi}^\mathrm{H}$ and $\boldsymbol{\phi}^\mathrm{W}$ using the most recent
approximations to $\tilde{\mathbf{X}}$ and $\tilde{\mathbf{X}}_t$ available; details are given in the appendix.

At the first iteration of each timestep the converged values from the previous
timestep are used as starting guesses for all variables, except for $\mathbf{X}$
which is approximated via linear extrapolation. The nonlinear iteration is
terminated when the maximum difference in position between successive iterations
relative to the distance travelled by the flagellum over the timestep falls
below $0.5\%$. Similarly, the auxiliary equation for the tension at $t+dt$ is
linearised as,
\begin{align}
\mathrm{Sp}^4(\lambda(1-\tilde{\mathbf{X}}_s\cdot\mathbf{X}_s) -\tilde{\mathbf{V}}_s\cdot\tilde{\mathbf{X}}_s) &=  \gamma T_{ss} - (\tilde{\mathbf{X}}_{ss}\cdot\tilde{\mathbf{X}}_{ss})T + 3\gamma \tilde{\mathbf{X}}_{sss}\cdot\mathbf{X}_{sss}+(1+3\gamma)\tilde{\mathbf{X}}_{ss}\cdot\mathbf{X}_{ssss} \nonumber \\
& +(\gamma +1)(\tilde{\mathbf{n}}_s\cdot\tilde{\mathbf{X}}_s)m_s + (\tilde{\mathbf{n}}_{ss}\cdot\tilde{\mathbf{X}}_s)m.  \label{eq:auxlinearised}
\end{align}
Each iteration requires the solution of a linear system for the unknown discrete
values of $\mathbf{X}(s_l,t_{n+1})$, $T(s_l,t_{n+1})$, $\mathbf{U}^\mathrm{H}$
and $\boldsymbol{\Omega}^\mathrm{H}$, where $l=0,\ldots,N_s$ denotes the spatial
grid coordinate and $n=0, 1, \ldots$ the timestep. We found that $N_s=160$ and
$200$ time steps per beat were sufficient to yield accurate results. The discrete
form of equations~\eqref{eq:elastohydro} and \eqref{eq:aux} provide $4(N_s+1) +
6 = 650$ linear equations, the additional $6$ equations arising from the
translational and angular velocity of the cell head. The nonlinear correction is
then a system of $3(N_s + N_h + N_b)$ linear equations, where $N_h$ and $N_b$
are the number of elements on the head and domain boundary respectively.

To implement the boundary conditions~\eqref{eq:distbc}, \eqref{eq:proxbc}, the force and moment on the head are \emph{a priori} unknown and need to be determined as part of the coupled problem. The force and moment are decomposed into a linear part, given by the grand resistance matrix associated with rigid body motion in the vicinity of the wall, and an additional subleading correction resulting from the influence of the flagellum. Following nondimensionalisation with the elasticity scalings, the force and moment on the head may then be expressed as,
\begin{equation}
\begin{pmatrix}\mathbf{F}^\mathrm{H} \\ \mathbf{M}^\mathrm{H}\end{pmatrix} = \mathrm{Sp}^4\left(\frac{\mu}{\xi_\perp}\right)\mathcal{R}\cdot \begin{pmatrix} \mathbf{U}^\mathrm{H} \\ \boldsymbol{\Omega}^\mathrm{H} \end{pmatrix}+\begin{pmatrix}\Delta\mathbf{F}^\mathrm{H} \\ \Delta\mathbf{M}^\mathrm{H}\end{pmatrix} \label{eq:forceCorrections}
\mbox{,}
\end{equation}
where $\Delta\mathbf{F}^\mathrm{H}$, $\Delta\mathbf{M}^\mathrm{H}$ are
corrections for the effect of the flagellum. The calculations of $\mathcal{R}$ and the corrections are described in the
appendix.

In summary, each timestep requires a number of iterations to solve the nonlinear
problem and each iteration involves the solution of a sparse linear system arising from the finite difference discretisation of the elastohydrodynamic equations. The `right hand side' terms arising from the nonlocal hydrodynamic correction
$\mathbf{V}$ and the nonlocal corrections to the force and moment balance
$\Delta \mathbf{F}^\mathrm{H}$, $\Delta\boldsymbol{\Omega}^\mathrm{H}$ require
the solution of a slender body theory-boundary integral hydrodynamic problem.
Calculation of the grand resistance matrix $\mathcal{R}$ requires the separate solution of a boundary integral problem with multiple right
hand sides to determine the force and moment resistances associated with the rigid body modes of the head and the wall interaction. The code is implemented in Fortran 90 (gfortran, GNU Compiler Collection); linear systems are equilibriated and solved by LU factorisation with
the LAPACK routines {\texttt{dgeequ}} and {\texttt{dgesv}} respectively, and the boundary
integrals over the sperm head are calculated with routines from BEMLIB \cite{pozrikidis2010}. A
typical run of $200$ beats with $500$ boundary elements required approximately 24 hours walltime on a
single core of a 2.2~GHz Intel Sandy Bridge E5-2660 node.

\section{Results}

\begin{figure}[tbp]	
\centering
\includegraphics{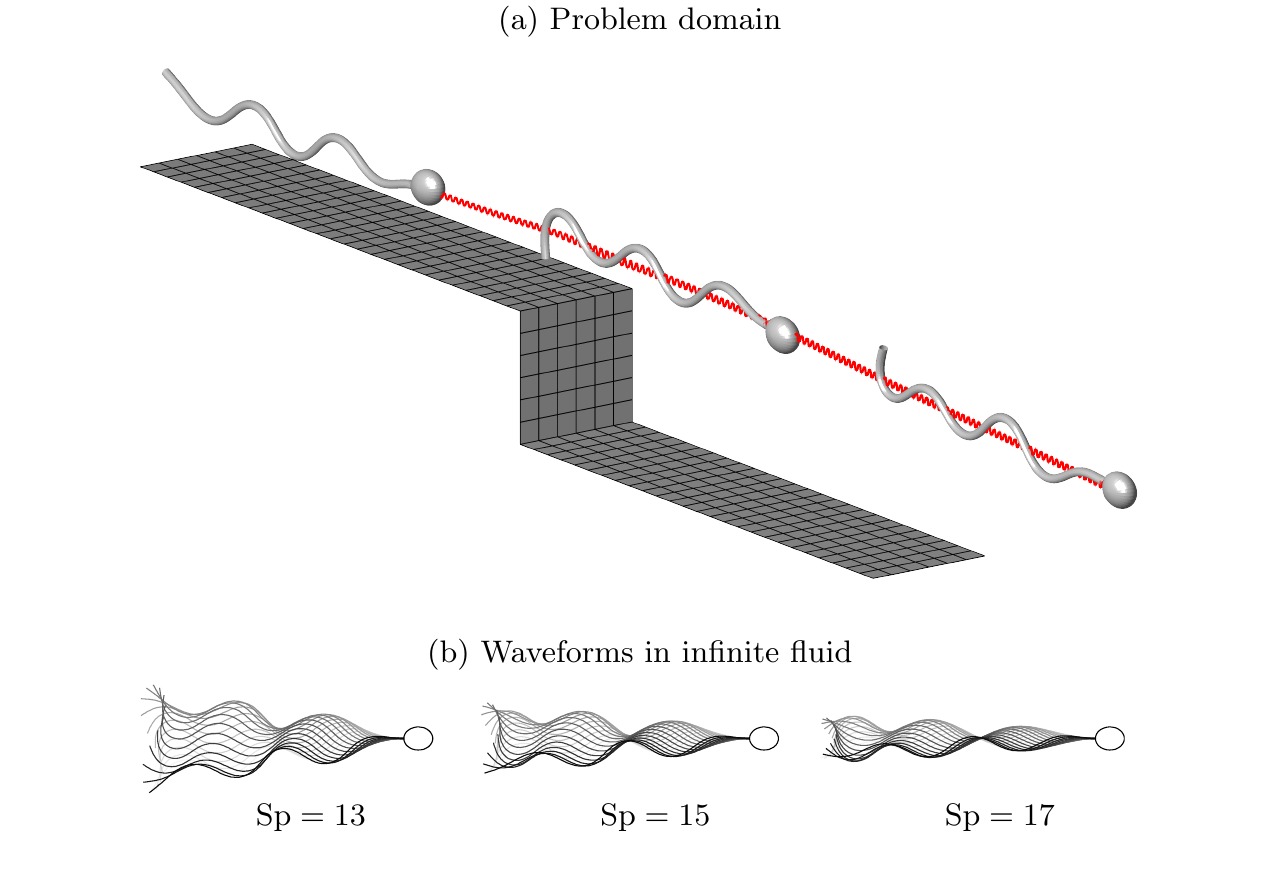}
\caption{Example results from nonlocal elastohydrodynamic simulation.
(a) Plot of the problem domain, including the boundary element meshes for the wall and swimmer, and a plot of the trajectory, computed at $\mathrm{Sp} = 13$. (b) Waveforms in infinite fluid of flagella driven by the same internal force at
different sperm numbers. The effect of increasing sperm number is to reduce cell
yaw and bending in the proximal end of the flagellum, as is observed in the
waveform of sperm in high viscosity medium \cite{smith2009bend}.}\label{fig1}
\end{figure}

The numerical scheme is applied to predict the trajectory of a sperm-like cell
swimming in an unbounded fluid at varying $\mathrm{Sp}$, over a `backstep' (the
latter being shown in figure~\ref{fig1}a), the limiting case of zero backstep
height being referred to as a `strip'. As in Gad\^{e}lha et
al.~\cite{gadelha2010nonlinear}, we consider planar waveform actuation, which is
appropriate for cells swimming through high viscosity fluids such as cervical
mucus \cite{smith2009bend}. The semi-axes of the ellipsoidal head, modelled
with the boundary element method, are $a_x = 0.05L$, $a_y = 0.03L$, $a_z =
0.04L$, which correspond to $5\times3\times4\,\mu m$ for a flagellum of length
$L=50$~$\mu$m. The swimmer is initially at rest, with a straight flagellum, and
a `soft start' is applied whereby the internal shear moment is initially low and
smoothly increases to its maximum, reaching 99\% after 5 beats. The sperm number
of a human gamete can be approximated by using bending stiffness $E \approx
5\times10^{-21}$~Nm$^2$, beat frequency $10$--$20$~Hz giving an angular
frequency $\omega \approx 100$~rad$/$s \cite{gaffney2011}.  Taking a flagellar
radius of $0.5$~$\mu$m, viscosity $\mu \approx 0.14$~Pa$\cdot$s (similar to
mucus analogue \cite{smith2009bend}) yields the normal resistance coefficient
$\xi_\perp\approx 0.503$ and sperm number $\mathrm{Sp}\approx 15.8$. Therefore,
we will consider a range of sperm numbers between $13$ and $17$, fixing the
magnitude of the internally generated shear-force $m_0=240$ and wavenumber
$k=6\pi$. The resulting waveforms are shown in figure \ref{fig1}(a).  As sperm
number increases, beat amplitude is suppressed, as is observed for sperm in high
viscosity medium \cite{smith2009bend}, leading to a reduction in side-to-side
yaw. All simulations in infinite fluid, i.e.\ with no nearby boundaries,
produced trajectories which were straight overall, once the within-beat yaw was
accounted for (data submitted to DRYAD
repository~\cite{montenegrojohnson2014}); flagellar waveforms for
$\mathrm{Sp}=13$, $15$ and $17$ are shown in figure~\ref{fig1}(b,c).

\begin{figure}[tbp]	
\centering
\includegraphics{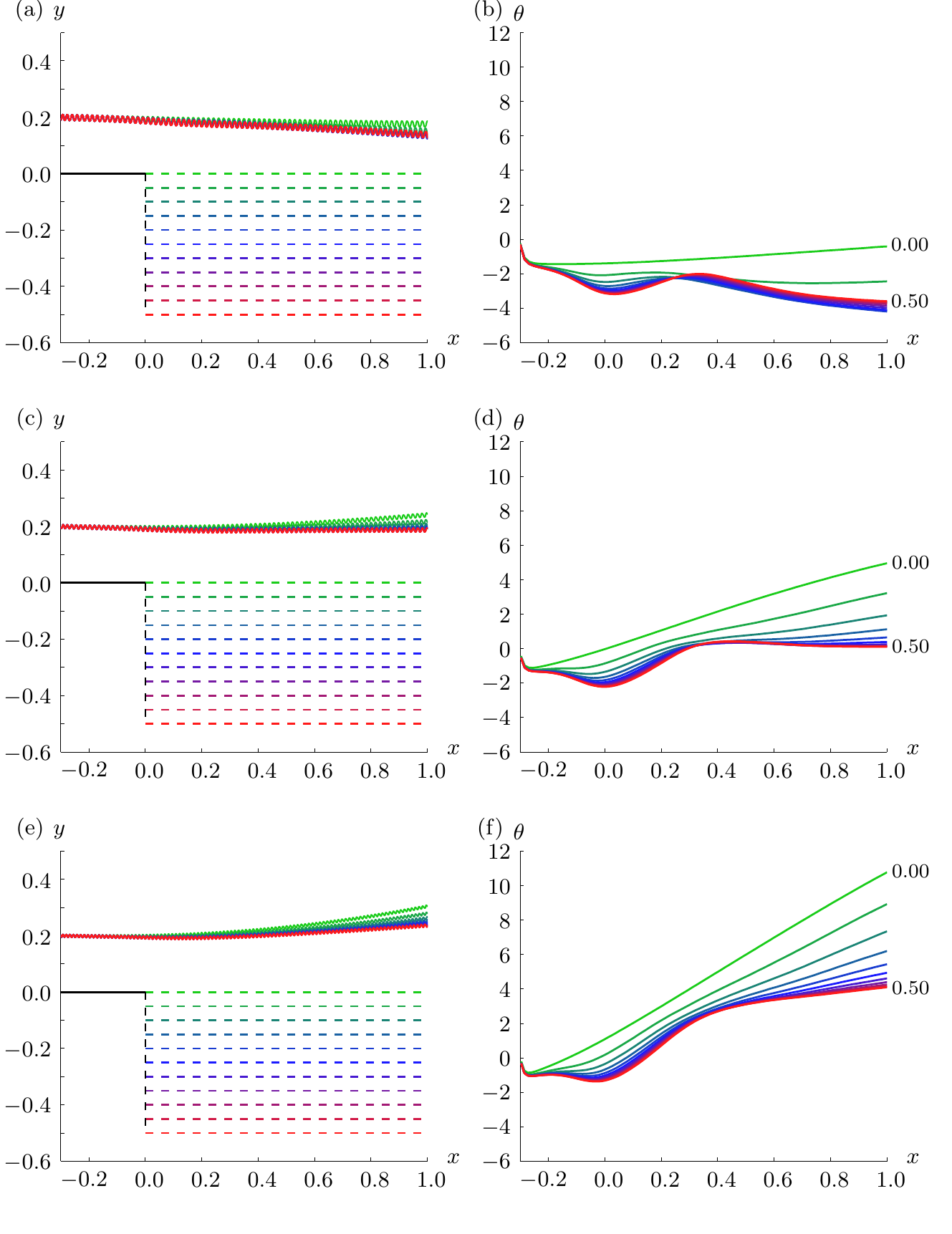}
\caption{Projected trajectories $(X(0,t),Y(0,t))$ and angles of trajectories
$\theta:=\arctan dY/dX(s=0)$ for different sperm numbers as a function of
changing the height of the backstep. (a, b) $\mathrm{Sp}=13$, (c, d)
$\mathrm{Sp}=15$, (e, f) $\mathrm{Sp}=17$. Colour corresponds to the backstep
heights shown in (a, c, e).}
\label{fig2}
\end{figure}

Figure~\ref{fig2} shows a planar projection of the trajectories $(X(0,t),Y(0,t))$, and the tangent angle $\theta:=\arctan (dY/dX(s=0))$ (in degrees) of those trajectories, of cells swimming over backsteps of varying height. The derivative $dY/dX$ is calculated numerically by sampling the trajectory at the temporal midpoint of each beat-cycle and taking centred differences. Colour indicates the trajectory over the backstep of the height in figure~\ref{fig2}(a, c, e) with green denoting $h = 0$ and red denoting $h = 0.5$. Simulations were performed over backsteps of height $h = 0.05,0.1,\dots,0.5$, and are displayed up to the time at which $X(0,t)\geqslant 1$.

The results in figure~\ref{fig2}(a, c, e) suggest that the backstep affects
swimmers at different sperm numbers differently, producing a range of scattering
angles. However, it is important in these results to factor out the effects of
the strip from the backstep. Taking the (lightest) green trajectory,
representing a strip, as a baseline comparison, it is evident that for all sperm
numbers the hydrodynamic effect of the backstep is to deflect the swimmer
downwards relative to a strip trajectory. Figure~\ref{fig2}(b, d, f) reveal that
this downward deflection is not smooth, rather there is a sharp bump at $x = 0$
where the head initially passes over the backstep, and a further bump at around
$x = 0.3$ where the effect of the step itself becomes subleading relative to
boundary interactions between the head and the lower wall.

Simulations were also performed comparing the effect of the backstep to a
`cliff' geometry, with the lower portion of the backstep missing (data
submitted~\cite{montenegrojohnson2014}). After passing the backstep, cells swam
straight as though in an infinite fluid, suggesting that the majority of the
angular deflection occurs due to interaction with the lower boundary; boundary
forces change suddenly over a step jump, and the cell acts as though it were
above a higher boundary. Additionally, simulations over a strip at $\mathrm{Sp}
= 13$ for different starting heights (data
submitted~\cite{montenegrojohnson2014}) showed that attraction to the surface
initially increased and then decreased as height above the surface increased,
which suggests that hydrodynamic boundary attraction is
responsible for the behaviour in figure~\ref{fig2}(a, b).

\begin{figure}[tbp]	
\centering
\includegraphics{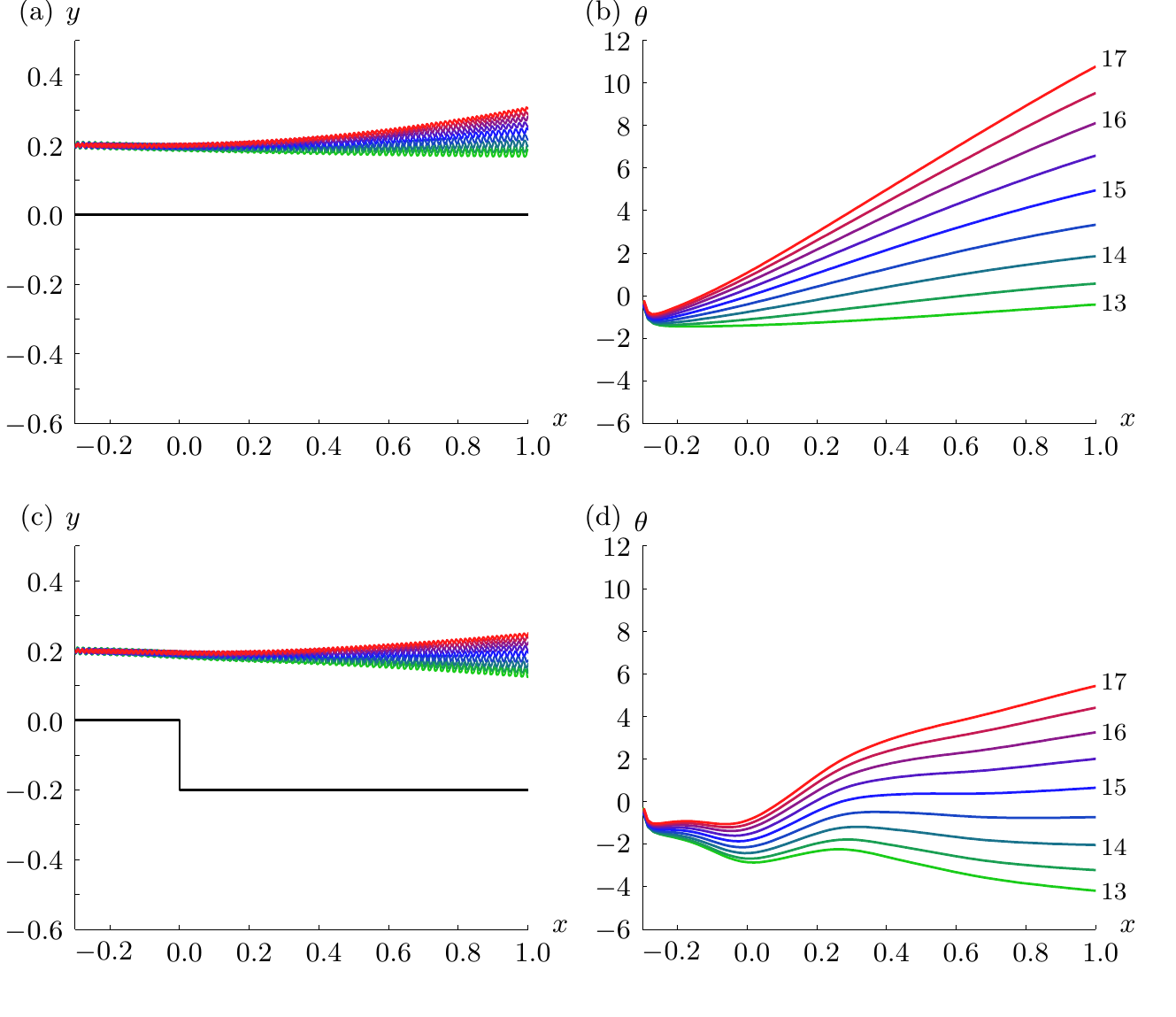}
\caption{Projected trajectories $(X(0,t),Y(0,t))$ and angles of trajectories
$\theta:=\arctan dY/dX(s=0)$ for varying sperm number over fixed geometry.(a, b) show
trajectories and angles with a `strip', i.e.\ zero backstep height, (c, d) with
backstep height $0.2L$. Colour is matched to sperm number, light green denoting
$\mathrm{Sp}=13$ and red denoting $\mathrm{Sp}=17$, intermediate colours moving
in increments of $0.5$.}
\label{fig3_a}
\end{figure}

\begin{figure}[tbp]	
\centering
\includegraphics{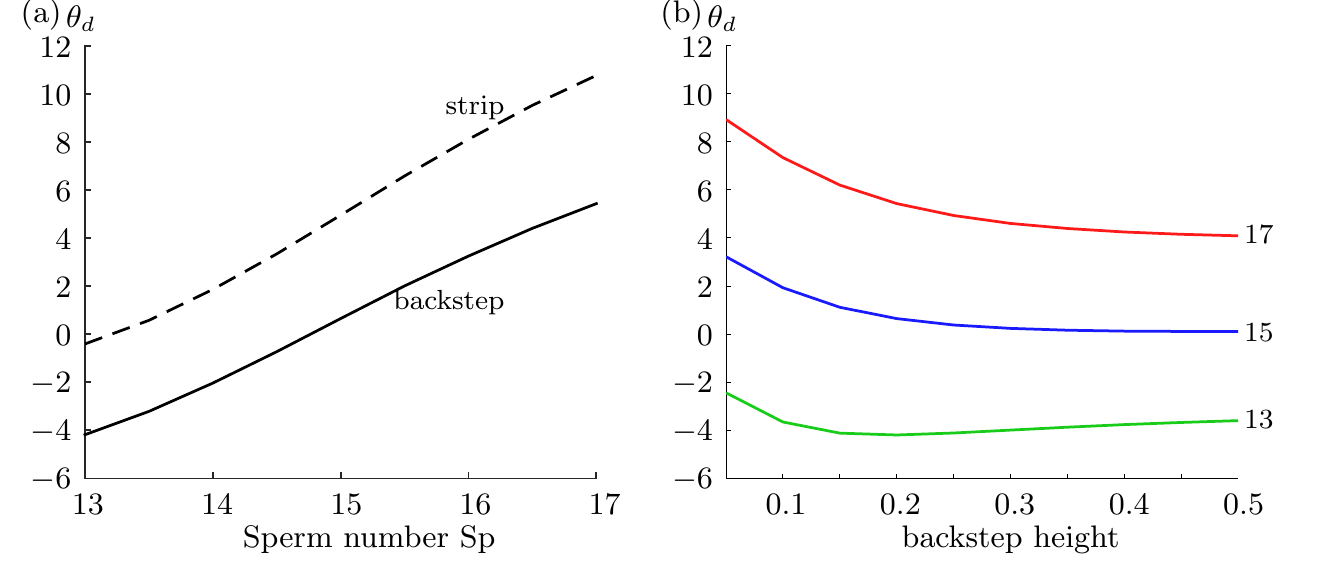}
\caption{The effect of the backstep on scattering, showing `final' deflection at
$x=1.0L$, for (a) strip and backstep ($h=0.2L$) and a range of sperm numbers,
(b) $\mathrm{Sp}=13, 15, 17$ and a range of backstep heights. Colour in (b)
denotes sperm number as indicated by labels on the right hand side.}
\label{fig3_b}
\end{figure}

Figure~\ref{fig3_a} shows the effect of varying sperm number over finer
increments for backstep height zero (a, b) and $h=0.2L$ (c, d), with results
summarised in figure~\ref{fig3_b}(a). Simulations were performed for
$\mathrm{Sp} = 13,13.5,\dots,17$ over both a strip geometry and a backstep of
height $h = 0.2L$, so that the sperm cells initially start $0.2L$ above the
surface, and then increase to around $0.4L$ after the backstep. In
figure~\ref{fig3_a}(a--d), colour is matched to increasing sperm number, so that
light green corresponds to $\mathrm{Sp} = 13$ and red to $\mathrm{Sp} = 17$.
Figure~\ref{fig3_a}(a, b) show for a sperm swimming over a strip, the boundary
repels the swimmer more at this close distance as sperm number is increased.
This effect is to be expected because increasing the sperm number increases the
relative strength of viscous to elastic forces, thus the effect of the boundary
is likely to be enhanced as $\mathrm{Sp}$ increases. The initial dip in
figure~\ref{fig3_a}(b) is an artifact of the numerical soft start of our system,
as the waveform emerges from a straight initial state.

Figure~\ref{fig3_a}(c, d) show a larger range of scattering angles than for fixed
sperm number over various backstep heights, of the order of $10^\circ$.
Furthermore, additional simulations (data
submitted~\cite{montenegrojohnson2014}) showed that this hydrodynamic deflection
was not sensitive to the phase of the waveform as it passed over the backstep,
in contrast to scattering due to contact forces (R.\ Goldstein, personal
communication, 2014). Figure~\ref{fig3_b}(a) shows the effects of changing sperm
number, giving the deflection for a strip, a backstep, and their difference. A
slight increase in the magnitude of this difference is observed as sperm number
is increased, owing to increased hydrodynamic interaction mediated by viscosity.

Figure~\ref{fig3_b}(b) summarises the effect of varying both backstep height and
sperm number simultaneously, quantified by the `final deflection angle'
$\theta_{\mathrm{d}}$, i.e.\ the value of $\theta$ for which $X=L$. At
$\mathrm{Sp}=13$ deflection is always negative, whereas for $\mathrm{Sp}=15$,
$17$ deflection is always positive. The relationship between $\theta_\mathrm{d}$
and $h$ is non-monotone at the lower sperm number but is monotonic in the higher
range. At $\mathrm{Sp}=13$, the deflection angle initially increases in
magnitude, then decreases after the maximum at around $h = 0.15L$. This riser
height corresponds to a distance of $0.35L$ between the cell and the boundary,
which is where boundary attraction is strongest at this sperm number. For
$\mathrm{Sp}=15$, $17$ the deflection angle decreases monotonically with
backstep height in the range we have considered. This effect likely occurs
because at these sperm numbers, the strip causes the cell to pitch away. However
in all cases increasing backstep height to $0.5L$ results in a plateau.

\begin{figure}[tbp]	
\centering
\includegraphics{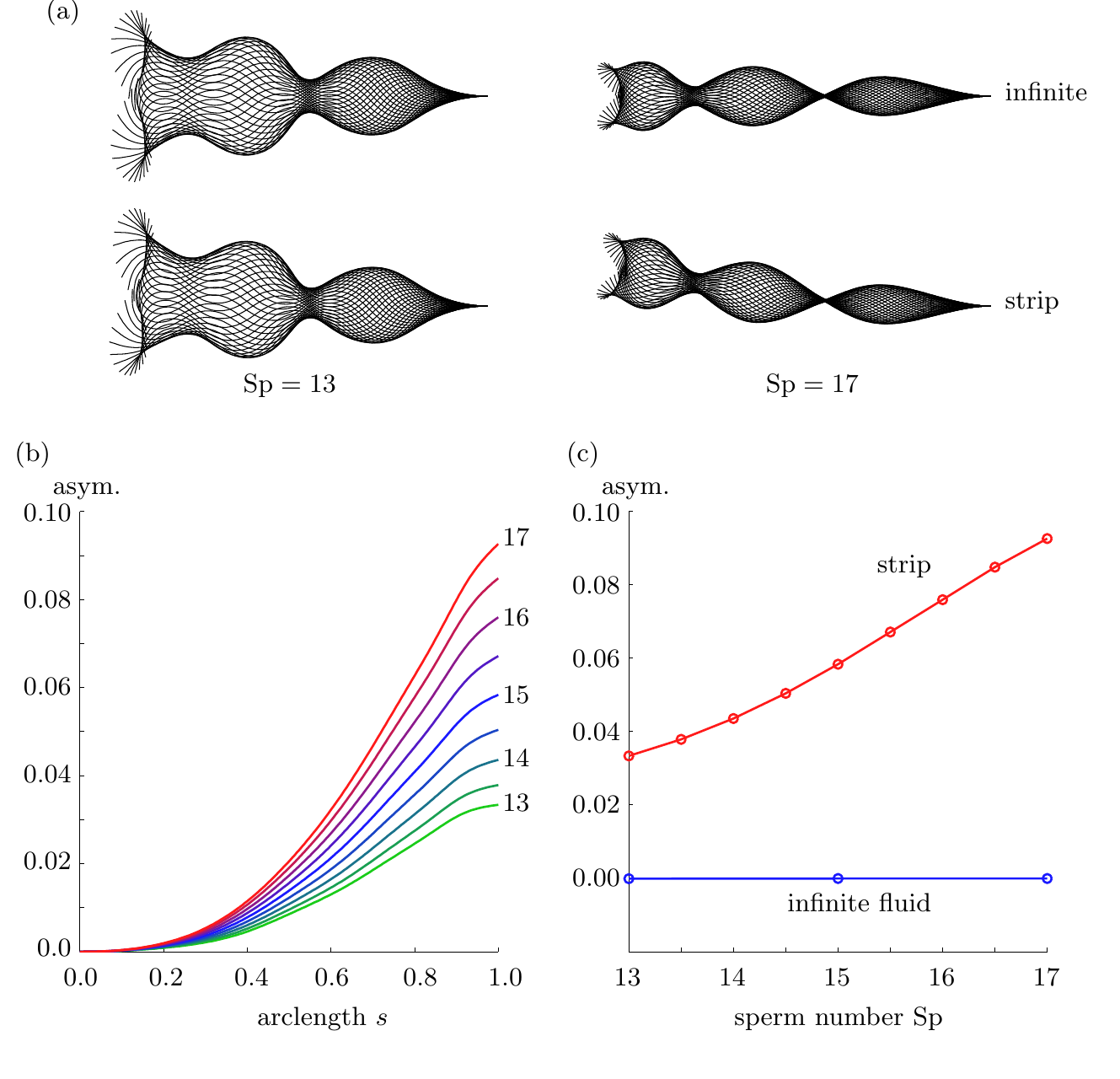}
\caption{Boundary-induced waveform asymmetry, increasing with sperm number. (a)
Symmetric waveforms of sperm flagella swimming in infinite fluid, and asymmetric
forms over the boundary strip for $\mathrm{Sp}=13$ and $\mathrm{Sp}=17$.  (b)
Flagellar asymmetry (defined in text) as a function of arclength, for a cell
swimming over a strip along the flagellum.  Colour in (b) is matched to sperm
number, light green denoting $\mathrm{Sp}=13$ and red denoting $\mathrm{Sp}=17$,
intermediate colours moving in increments of $0.5$. (c) Asymmetry at $s=1$ at
the endpoint of the flagellum, as a function of sperm number $\mathrm{Sp}$,
showing an increase in asymmetry with sperm number for the strip (red), and
negligible asymmetry for the infinite fluid case (blue).}
\label{fig4}
\end{figure}

The effects of the backstep on the waveform are summarised in figure~\ref{fig4},
which show the waveform shape with and without the boundary, and quantitative
measures of the asymmetry of the waveform. Recall that the flagellar actuation
is symmetric; waveform asymmetry is produced due to increased hydrodynamic drag
arising from proximity to the wall \cite{katz1975} affecting closer portions of
the flagellum more than further portions. Figure~\ref{fig4}(a) shows waveforms
at sperm number $\mathrm{Sp} = 13$, $17$ in infinite fluid as well as over a
strip. In infinite fluid, the waveform is symmetrical for all sperm numbers
considered, while the presence of a boundary gives rise to a waveform asymmetry
that increases with $\mathrm{Sp}$.

`Asymmetry' is quantified by sampling the flagellar wave every $41$ numerical
timesteps (relative to a beat cycle of $200$ timesteps), projecting into the
body frame, and calculating the average lateral position relative to the body
frame centreline over a fixed period, in this case beats $82$--$90$. This
quantity is plotted as a function of arclength in figure~\ref{fig4}(b); its
distal ($s=1$) value is plotted in figure~\ref{fig4}(c).

Figure~\ref{fig4}(b) plots asymmetry versus arclength for sperm numbers in the
range $13$--$17$, the effect being largest at higher sperm number. The asymmetry
at the tip of the flagellum for a strip versus no boundary is shown in
figure~\ref{fig4}(c) as a function of sperm number.

\section{Discussion}
A numerical method for simulating the swimming of monoflagellate cells over
geometric features was presented and applied to model sperm interacting with
microchannel backstep feature. The scheme incorporates nonlocal hydrodynamics
with large-amplitude active filament mechanics. We believe this method to be the
simplest generalisation of previous work that is capable of taking into account
nonlocal hydrodynamic interaction geometrical features. The linearity of the
Stokes flow equations entails that the largest error in our method arises from
the LGL slender body theory, which is at worst on the order of the square root
of the slenderness ratio. Accuracy of the method of regularised stokeslets is on
the order of the regularisation parameter near the boundary, and its square far
from the boundary where the swimmer is located.  Future work may consider
boundary integral modelling of the flagellum also, however we do not expect that
this would qualitatively change swimmer trajectories.

The interaction between the cell and the lower boundary involves the competing
effects of asymmetric hydrodynamic forces leading to waveform asymmetry and
boundary repulsion, and the pitching behaviour associated with swimmer/boundary
interaction \cite{smith2009human}. At lower sperm number and at greater
distances from the boundary, waveform asymmetry is smaller, and the cell pitches
towards the boundary. At higher sperm number and closer distances from the
boundary, waveform asymmetry is larger and the cell pitches away. The effect of
the backstep is a sudden drop in the lower boundary, which changes the relative
importance of these effects; waveform asymmetry is reduced relative to
hydrodynamic attraction, and the net result is a deflection towards the lower
boundary after the backstep relative to the expected trajectory over a strip
(figure~\ref{fig2}).


Analysing sperm scattering over a backstep, we found that hydrodynamic effects
may be comparable in magnitude in the relatively high viscosity range considered
to the contact interactions found experimentally by Kantsler et
al.~\cite{kantsler2013ciliary}.  A transition is predicted from scattering
towards the backstep at lower viscosity to scattering away from the backstep at
higher viscosity. Qualitatively this behaviour is similar to the
temperature-related transition in Kantsler et al.'s observations (with lower
temperature corresponding to higher viscosity); the correspondence is not exact
however, with Kantsler et al.'s observations being carried out with bull sperm
in low viscosity buffer, and with cells exhibiting very close interaction with
the boundary, compared with our longer range interactions and sperm number
representative of human cells in mucus analogue that we chose to focus on in the
present study. Clearly integrating both surface interactions and hydrodynamics
will be necessary to develop a comprehensive model, particularly at higher sperm
number/viscosity.

The role of hydrodynamic interactions in determining surface attraction and more
complex effects associated with boundary features continues to receive
significant theoretical attention and is stimulating novel mathematical
approaches \cite{crowdy2011,davis2012,spagnolie2012,crowdy2013,lopez2014}.
Viscous interactions of course become increasingly important in high viscosity
fluids such as mucus and laboratory analogues. Kantsler et
al.~\cite{kantsler2013ciliary} noted the need to take both elastic and steric
interactions into account; modelling very short length scale or contact
interactions, with either glass, epithelium, cumulus, or even ciliated surfaces,
and their effect on the flagellar wave, is a topic of importance, though
numerical simulation requires taking account of the rapidly varying hydrodynamic
force and electrostatic interactions as the swimmer approaches these boundaries.
We hope that the numerically implicit method, potentially also combined with
adaptive refinement of the boundary element meshes, will enable accurately
resolved simulation of sperm-like swimmers in very near surface-contact in
future work. Other valuable methods for modelling three-dimensional sperm
motility and elastic-fluid interaction include models based exclusively on
regularised stokeslets \cite{nguyen2011,cortez2012} and techniques such as
stochastic rotation dynamics \cite{elgeti2013}.

Whilst we have used our model to examine a swimmer representative of human
sperm, the approach is applicable to a much wider range of eukaryotic cells,
including the sperm of other species and, with a slight reworking of the head
boundary condition, biflagellate organisms such as the green alga \emph{Chlamydomonas}.
These species are of particular interest as they have been used as models
for flagellar synchronisation \cite{wan2014lag} and are relevant to energy-producing bioreactors \cite{bees2014mathematics}. For these systems, the model may also be extended to include a nonlocal hydrodynamic contribution from other swimmers. Larger swimming organisms, such as \emph{C. elegans}, have also been shown to be significantly affected by interactions with a structured microenvironment \cite{park2008,majmudar2012}.

Another application are is the design and optimisation of biomimetic artificial
microswimmers (see for example refs.~\cite{dreyfus2005,espinosa2013}). Because the model includes internal periodic actuation via
prescribed bending moments, it might be used to optimise actuation for various
purposes such as forward progress, subject to constraints such as fixed mechanical energy. Furthermore, the inclusion of geometrical boundary features and the use of sperm number allows such optimisation to be
tailored to specific environments. The elastohydrodynamic model can additionally be
used to solve the inverse problem of estimating internal moments from observed flagellar data, potentially allowing us to examine how nature has optimised swimming in various
environments and informing truly biomimetic design.

Despite the linearity of the Stokes flow equations, the interaction of sperm with their microenvironment presents a subtle nonlinear mechanics problem. Sperm scattering depends nonlinearly on the ratio between viscous and elastic forces, with even a simple backstep feature producing attractive or repulsive scattering of cells depending on parameter values. These scattering effects may be valuable in sorting cells in microdevices, in addition to giving insight into the complexity of how sperm interact with their microenvironment. The combination of mechanical models and experiment will provide the best way to understand and exploit these effects for biomedical applications.

\section*{Data accessibility}
Trajectory and waveform data supporting the findings in this paper have been
submitted to DRYAD; see reference item~\cite{montenegrojohnson2014}.

\section*{Competing interests}
None.

\section*{Authors' contributions}
TDMJ designed the research, implemented algorithms, analysed results and co-wrote the manuscript. HG designed the research and contributed to the writing of the manuscript. DJS designed the research, helped with algorithmic implementation, co-wrote the manuscript and supervised the project.

\section*{Acknowledgements}
The authors thank Dr Eamonn Gaffney (University of Oxford), Prof.~Ray Goldstein (University of Cambridge), Dr Vasily Kantsler, Dr Petr Denissenko (University of Warwick), Prof.~John Blake (University of Birmingham) and Dr Jackson Kirkman-Brown (University of Birmingham/Birmingham Women's NHS Foundation Trust) for continuing valuable discussions.

\section*{Funding}
This work was supported by EPSRC grant EP/K007637/1 to DJS, supporting TDMJ.
TDMJ is supported by a Royal Commission for the Exhibition of 1851 Research
Fellowship. HG acknowledges a Hooke Fellowship held at the University of Oxford.
The computations described in this paper were performed using the University of
Birmingham's BlueBEAR HPC service, which was purchased through SRIF-3 funds. See
http://www.bear.bham.ac.uk for more details.

\section*{Appendix: calculation of hydrodynamic terms}
Following nondimensionalisation, the hydrodynamic model yields the following equation for the dimensionless fluid velocity away from the flagellum,
\begin{align}
\mathrm{Sp}^4\mathbf{U}(\mathbf{x}) &= \frac{\xi_\perp}{\mu}\left(\int_0^1\boldsymbol{\mathsf{S}}(\mathbf{x},\mathbf{X}(s,t))\cdot\mathbf{f}_\mathrm{vis}(s,t) ds + \iint_{H(t)}\boldsymbol{\mathsf{S}}(\mathbf{x},\mathbf{Y})\cdot\boldsymbol{\phi}^\mathrm{H}(\mathbf{Y},t) dS_{\mathbf{Y}}\right.\nonumber \\
&\left.+\iint_W\boldsymbol{\mathsf{S}}^\epsilon(\mathbf{x},\mathbf{Y})\cdot\boldsymbol{\phi}^\mathrm{W}(\mathbf{Y},t) dS_{\mathbf{Y}}\right).
\end{align}%
The nonlocal contribution to the velocity $\mathbf{V}$ on the slender body is similarly given by,
\begin{align}
\mathrm{Sp}^4\mathbf{V}(\mathbf{X}(s_0,t)) &=  \frac{\xi_\perp}{\mu}\left(\int_{|s-s_0|>q}\boldsymbol{\mathsf{S}}(\mathbf{X}(s_0,t),\mathbf{X}(s,t))\cdot\mathbf{f}_\mathrm{vis}(s,t) ds \right. \nonumber \\
& + \iint_{H(t)}\boldsymbol{\mathsf{S}}(\mathbf{X}(s_0,t),\mathbf{Y})\cdot\boldsymbol{\phi}^\mathrm{H}(\mathbf{Y},t) dS_{\mathbf{Y}}\nonumber \\
&\left.+\iint_W\boldsymbol{\mathsf{S}}^\epsilon(\mathbf{X}(s_0,t),\mathbf{Y})\cdot\boldsymbol{\phi}^\mathrm{W}(\mathbf{Y},t) dS_{\mathbf{Y}}\right).\label{eq:nonloc}
\end{align}%

At each step of the iterative solution to the nonlinear problem, the collocation code solves the integral equation,
\begin{equation}
\mathrm{Sp}^4\begin{pmatrix}\mathbf{X}_t\\\mathbf{U}^\mathrm{H}+\boldsymbol{\Omega}^\mathrm{H}\wedge(\mathbf{Y}^\mathrm{H}-\mathbf{X}^\mathrm{c})\\0\end{pmatrix}
=
\begin{pmatrix}
-(\boldsymbol{\mathsf{I}}+(\gamma-1)\hat{\mathbf{s}}\hat{\mathbf{s}})\cdot \mathbf{f}_{\mathrm{vis}}+\mathrm{Sp}^4\mathbf{V}[\mathbf{f}_{\mathrm{vis}},\boldsymbol{\phi}^\mathrm{H},\boldsymbol{\phi}^\mathrm{W}](\mathbf{X})
\\
\mathrm{Sp}^4\mathbf{U}[\mathbf{f}_{\mathrm{vis}},\boldsymbol{\phi}^\mathrm{H},\boldsymbol{\phi}^\mathrm{W}](\mathbf{Y}^\mathrm{H})
\\
\mathrm{Sp}^4\mathbf{U}[\mathbf{f}_{\mathrm{vis}},\boldsymbol{\phi}^\mathrm{H},\boldsymbol{\phi}^\mathrm{W}](\mathbf{Y}^\mathrm{W})
\end{pmatrix}\mbox{,}
\end{equation}
for the unknown hydrodynamic force per unit length $\mathbf{f}_{\mathrm{vis}}$ and unknown stresses $\boldsymbol{\phi}^\mathrm{H}$, $\boldsymbol{\phi}^\mathrm{W}$.

The collocation code discretises the flagellum with $160$ elements, with the nonlocal contribution
to the LGL slender body theory computed by the midpoint rule with constant force
per unit length over each element. The force per unit area on the ellipsoidal head
of $32$ mesh elements is calculated using routines from BEMLIB \cite{pozrikidis2010} with $20$ point
Gauss-Legendre quadrature as described in detail in the appendix. The wall boundary is discretised into elements of width
$0.075L$, using regularised stokeslets with $\epsilon = 0.01L$. Integration is performed with repeated Gauss-Legendre quadrature with $4\times 4$ points per element for the near-singular wall integrals, and a $2\times 2$ point rule elsewhere.

To implement the boundary conditions~\eqref{eq:distbc}, \eqref{eq:proxbc}, Gad\^{e}lha et al.~\cite{gadelha2010nonlinear} approximated the force and moment on the head by a grand resistance matrix \cite{pozrikidis2010} multiplying the velocity and angular velocity. In dimensional variables, the grand resistance matrix expresses the force and moment on a moving rigid body as
\[
\begin{pmatrix}\mathbf{F} \\ \mathbf{M}\end{pmatrix} = \mathcal{R}\cdot \begin{pmatrix} \mathbf{U} \\ \boldsymbol{\Omega} \end{pmatrix} , \quad \mbox{where} \quad \mathcal{R} =  \begin{pmatrix} \quad &  \mathcal{R}^\mathrm{F} &  \\ & \mathcal{R}^\mathrm{M} & \end{pmatrix}.
\]
The blocks $\mathcal{R}^\mathrm{F}$ and $\mathcal{R}^\mathrm{M}$ are $3\times 6$ matrices yielding the force and moment terms respectively. For example, a sphere of radius $a$ in the absence of hydrodynamic interactions would have dimensionless grand resistance matrix given by
\begin{equation}
\mathcal{R}^\mathrm{F}=\begin{pmatrix} \quad & -6\pi \mu a\boldsymbol{\mathsf{I}} &  0 & \quad \end{pmatrix}, \quad \mathcal{R}^\mathrm{M}=\begin{pmatrix}\quad & 0 &  -8\pi \mu a^3\boldsymbol{\mathsf{I}}  & \quad \end{pmatrix}.
\end{equation}
This approach is convenient because the linearity of the relationship means that the head velocity $\mathbf{U}^\mathrm{H}$ and angular velocity $\boldsymbol{\Omega}^\mathrm{H}$ can be dealt with in the implicit formulation as unknowns in the linear algebra problem. To generalise to a nonlocal hydrodynamic model taking into account the effect of the flagellum and nearby boundary, the force and moment will be decomposed as consisting of part which is linear in velocity and angular velocity via resistance matrices and a remaining contribution from the flagellum. The matrices $\mathcal{R}^\mathrm{F}$ and $\mathcal{R}^\mathrm{M}$ are determined via the boundary integral method, taking into account the potentially highly significant effect of the wall feature, but not the subleading effect of the flagellum, which is accounted for as a correction, as described below.

Elastic scalings are used to nondimensionalise all forces and moments, i.e.\ $E/L^2, E/L$ for $\mathbf{F}^\mathrm{H}$ and $\mathbf{M}^\mathrm{H}$ respectively, with $E/L^3$ for force per unit length $\mathbf{f}_{\mathrm{vis}}$ and $E/L^4$ for stress $\boldsymbol{\phi}^\mathrm{H}, \boldsymbol{\phi}^\mathrm{W}$. The additional corrections $\Delta\mathbf{F}^\mathrm{H}$ and $\Delta\boldsymbol{M}^\mathrm{H}$ referred to in equation~\eqref{eq:forceCorrections} are determined as part of the iterative process by performing a slender body/boundary integral calculation of $\tilde{\mathbf{f}}_{\mathrm{vis}}$, $\tilde{\boldsymbol{\phi}}^H$ and $\tilde{\boldsymbol{\phi}}^W$ with the most recent approximation to $\tilde{\mathbf{X}}$ available, yielding in dimensionless variables,
\begin{equation}
\tilde{\mathbf{F}}^\mathrm{H}=\iint_{H(t)}\tilde{\boldsymbol{\phi}}^\mathrm{H}\,dS, \quad \tilde{\mathbf{M}}^\mathrm{H}=\iint_{H(t)}(\tilde{\mathbf{Y}}-\tilde{\mathbf{X}}^\mathrm{c})\wedge\tilde{\boldsymbol{\phi}}^\mathrm{H}\,dS_{\mathbf{Y}} \mbox{,}
\end{equation}
where $\tilde{\mathbf{X}}^\mathrm{c}$ is the head centroid. Using also the most recent iterates for $\tilde{\mathbf{U}}^\mathrm{H}$ and $\tilde{\boldsymbol{\Omega}}^\mathrm{H}$, the corrections are then given by,
\begin{equation}
\Delta\mathbf{F}^\mathrm{H} = \tilde{\mathbf{F}}^\mathrm{H} - \mathrm{Sp}^4\left(\frac{\mu}{\xi_\perp}\right)\mathcal{R}^\mathrm{F} \cdot \begin{pmatrix} \tilde{\mathbf{U}}^\mathrm{H} \\ \tilde{\boldsymbol{\Omega}}^\mathrm{H} \end{pmatrix} \mbox{,} \quad
\Delta\boldsymbol{M}^\mathrm{H} =\tilde{\mathbf{M}}^\mathrm{H} - \mathrm{Sp}^4\left(\frac{\mu}{\xi_\perp}\right)\mathcal{R}^\mathrm{M} \cdot \begin{pmatrix} \tilde{\mathbf{U}}^\mathrm{H} \\ \tilde{\boldsymbol{\Omega}}^\mathrm{H} \end{pmatrix}.
\end{equation}
These corrections appear on the right hand side of the linear system.

\bibliographystyle{vancouver}
\bibliography{nl_refs}

\vspace{2cm}

\begin{center}
\textbf{Short title for page headings:}
\emph{Scattering of sperm by a microchannel feature}
\end{center}


\end{document}